\documentclass[aps,pre,print,groupedaddress]{revtex4-1}
\usepackage{graphicx}  
\usepackage{dcolumn}  
\usepackage{bm}           
\usepackage{amsmath}
\usepackage{amsfonts}
\usepackage{amssymb}
\usepackage{color}
\usepackage[normalem]{ulem}
\usepackage{ wasysym }
\usepackage{dsfont}
\usepackage{mathtools}
\usepackage{amsthm}

\usepackage{array}
\newcolumntype{L}[1]{>{\raggedright\let\newline\\\arraybackslash\hspace{0pt}}m{#1}}
\newcolumntype{C}[1]{>{\centering\let\newline\\\arraybackslash\hspace{0pt}}m{#1}}
\newcolumntype{R}[1]{>{\raggedleft\let\newline\\\arraybackslash\hspace{0pt}}m{#1}}

\newtagform{AArabic}[]()

\usepackage{todonotes}

\theoremstyle{definition}

\usepackage{mdframed}
\newmdtheoremenv[style=MyFrame]{teorema}{Result}
\newmdtheoremenv[style=MyFrame]{estrategia}{Strategy}
\newmdtheoremenv[style=MyFrame]{nota}{Remark}
\newmdtheoremenv[style=MyFrame]{consecuencia}{Consequence}
\newmdtheoremenv[style=MyFrame]{definicion}{Definition}

\mdfdefinestyle{MyFrame}{%
    linewidth=0.4mm,
    linecolor=black,
    outerlinewidth=2pt,
    roundcorner=20pt,
    innertopmargin=4pt,
    innerbottommargin=4pt,
    innerrightmargin=4pt,
    innerleftmargin=4pt,
        leftmargin = 4pt,
        rightmargin = 4pt,
        backgroundcolor=gray!50!white
        }

\begin{document}

\title{Turing patterns on a two-component isotropic growing system.\\ Part 2: Conditions based on a potential function for exponential growth/shrinkage}

\author{Aldo Ledesma-Dur\'an}
\email{aldo\_ledesma@xanum.uam.mx}
\affiliation{UAM-Iztapalapa}

\date{\today}

\begin{abstract}
We propose conditions for the emergence of Turing patterns in a domain that changes in size by homogeneous growth/shrinkage. These conditions to determine the bifurcation are based on considering the geometric change of a potential function whose evolution determines the stability of the trajectories of all the Fourier modes of the perturbation. For this part of the work we consider the situation where the homogeneous state of the system are constant concentrations close to its stationary value, as occurs for exponential growth/decrease. This proposal recovers the traditional Turing conditions for two-component systems in a fixed domain and is corroborated against numerical simulations of increasing/decreasing domains of the Brusselator reaction system. The simulations carried out allowed us to understand some characteristics of the pattern related to the evolution of its amplitude and wavenumber and allow us to anticipate which features strictly depend on its temporal evolution.
\end{abstract}

\pacs{}

\maketitle

%
%
%
%
%
%
%
%
%
%
%
%

\section{Presentation}\label{sec:presentation}

Turing patterns in reaction-diffusion systems where the domain changes size present different aspects to those of a fixed domain, the most important being that the shape of the pattern at a specific time depends crucially on its past history \cite{krause2019influence,klika2017history}. It is believed that this phenomenon, which can be understood as a type of hysteresis, is probably related to persistence, \emph{i.e.}, the ability of a dissipative structure to maintain its current wavenumber, even when there is another state with a wavenumber  that can be more stable, \emph{i.e.}, more resistant to sideband disturbances \cite{ledesma2020eckhaus}. To test this possibility for any reaction-diffusion system or type of growth, a non-linear approximation to the solution of the reaction diffusion dilution (RDD) system near the Turing bifurcation  is required. However, in the case of a domain that changes with time, this analysis is impractical since it has not been conclusively resolved, even from the linear approach, how to find the Turing bifurcation in an increasing domain. Some important approaches to this respect are those in \cite{madzvamuse2010stability,van2021turing}.

The strategy followed in this work to find the Turing bifurcation is to consider the changes in the structure of the phase plane in order to find a  potential function for the Fourier modes perturbations. From this potential function, it is expected that all the trajectories decrease to stable point in absence of diffusion, and some become unstable saddles for some wavenumber when diffusion is turned on. We will show that the observance of the geometrical structure changes of the potential function allow one to establish some hypotheses of where Turing pattern can emerge. These hypotheses will be tested with specific numerical simulations of the Brusselator RDD system using finite differences method in one-dimensional reaction diffusion system with different types of homogeneous growing/shrinking.

\subsection{Summary of Part1: Homogeneous state and perturbations}\label{sec:antecedents}

Consider a reaction-diffusion process in a domain that grows in size $l(t)$. If growth occurs homogeneously, the relationship between the real and computational domain can be written as $ x =x_0 +l(t) \xi$ with $\xi \in [0, 1]$, where $\xi$ the fixed coordinate and $x$ the actual coordinate. If $\mathbf{c}$ represents the vector of concentrations, $\mathds{D} $ its diffusion matrix, and $\mathbf{f}( \mathbf{c})$ the vector of chemical reactions, we prove in Part 1 of this series that the RDD equation describing the dynamics \cite{crampin1999reaction} obeys

\begin{equation}\label{eq:system}
 \frac{\partial  \mathbf{c}}{ \partial t}  +   \frac{\dot{l}(t)}{l(t)} \mathbf{c}(\xi,t) = \frac{1}{l^2(t)}   \mathds{D}  \frac{ \partial^2  \mathbf{c}}{ \partial \xi^2}  +\mathbf{f}( \mathbf{c}).
 \end{equation} 
To exemplify some differences in the occurrence of patterns in fixed and increasing domains, in this work we will consider in our simulations the Brusselator given by \cite{pena2001stability} 

\begin{equation}\label{eq:bruselas1}
 \mathbf{f}(\mathbf{c})=(A-Bc_u-c_u +c_u^2 c_v, Bc_u -c_u^2 c_v)^T.
\end{equation}


The homogeneous state $\mathbf{c}_s(t)$, \emph{i.e.} that related to the zero Fourier mode of the solution of \eqref{eq:system} with no spatial contribution obeys

\begin{equation}\label{eq:homogeneous}
 \frac{\partial   \mathbf{c}_s}{ \partial t} +\frac{l'(t)}{l(t)} \mathbf{c}_s=\mathbf{f}(\mathbf{c}_s),
\end{equation} 
and presents very discernible features in each case. For the non-equilibrium Brusselator system, the dilution term can induce the following scenarios: 1) the homogeneous solution for large times can be different from the fixed point solution $\mathbf{c}_0$, \emph{i.e} that where  both reaction rates are null (for exponential growing/shrinking); 2) the homogeneous state can oscillate around the fixed-point concentration (for sinusoidal variation); and 3 ) the homogeneous state can tend very slowly to the fixed point concentration or diverge when the domain shrinks too much (for linear and quadratic shrinking). In contrast, for systems whose fixed point is the origin, all the types of growing lead the concentrations to the same final steady state defined by the fixed point at the origin. 

Under appropriate approximations related to 1) slow domain variation, 2) safe distance from bifurcations and 3) smallness of non-linear terms, we prove with comparisons with numerical solutions that a good approximation for the homogeneous state in \eqref{eq:homogeneous} is given by

\begin{equation}\label{eq:homoanalitic}
 \mathbf{c}_s(t)= \mathbf{c}_0+  \frac{l(0)}{l(t)} \mathds{P}  e^{\boldsymbol{\Lambda} t} \mathds{P}^{-1}  [ \mathbf{c}_s(0)- \mathbf{c}_0]-
 \mathds{P} \frac{e^{\boldsymbol{\Lambda} t}}{l(t)} \left(\int \limits_0^t  l'(t') e^{-\boldsymbol{\Lambda} t'}dt' \right)\mathds{P}^{-1}  \mathbf{c}_0.
\end{equation}
In this equation $\boldsymbol{\Lambda}$ is the diagonal matrix of eigenvalues of the Jacobian $\mathds{J}\equiv \frac{\partial f}{\partial \mathbf{c}}(\mathbf{c}_0)$, $\mathds{P}$ its modal matrix and $\mathbf{c}_0$ the constant fixed point. We prove how this approximation correctly describes the main features of the homogeneous state for exponential, linear, quadratic and sinusoidal domain variations.

We also show that for certain types of domain variation,  the concentration changes are small and therefore, a representative constant value $\mathbf{C}_0$  can approximate the homogeneous state $\mathbf{c}_s(t)$ on a time interval between $t_i$ and $t_{f}$. One form of this representative value is given by its direct  temporal average :

 \begin{equation}\label{eq:representative}
 \mathbf{C}_0  = \mathbf{c}_0 +\frac{1}{t_f-t_i}  \left( \int \limits_{t_i}^{t_f }\boldsymbol{\delta C}(t) \, dt \right) \mathbf{c}_0,
\end{equation} 
The explicit deviation $\boldsymbol{\delta C}(t)$ was calculated explicitly in Part 1 of this paper for exponential, linear, quadratic, and sinusoidal growth functions by averaging the equation \eqref{eq:homoanalitic}. We prove with comparisons with the numerical solutions that this equation approximates $\mathbf{c}_s(t)$ with a low error for slow growth/decrease rates, once a transient time has passed, and up to an increase/decrease time interval of ten times the size of the original domain.

This distinction between fixed point concentration $\mathbf{c}_0$, time dependent homogeneous state $\mathbf{c}_s(t)$ and representative and constant concentration $\mathbf{C}_0$, is relevant for non-equilibrium systems like the Brusselator. For systems like the BVAM, in the steady state, these three concentrations represent the same point: the origin \cite{leppanen2004computational,toole2013turing}.

The importance of this summary here lies in the effect that the homogeneous state has on the perturbations. In Part 1 we prove that the perturbations $\boldsymbol{\zeta}$ of the system \eqref{eq:system}, to first order obey 
\begin{equation}\label{eq:perturbations}
 \frac{\partial \boldsymbol{\zeta}}{ \partial t}   +\frac{\dot{l}(t)}{l(t)}\boldsymbol{\zeta} = \frac{1}{l^2(t)}   \mathds{D}  \frac{ \partial^2  \boldsymbol{\zeta}}{ \partial \xi^2}  + \frac{\partial \mathbf{f}}{\partial \mathbf{c}}(\mathbf{c}_s) \boldsymbol{\zeta}.
\end{equation} 
Therefore, the evaluation of the last term in general depends explicitly on the time-dependent homogeneous state. However, it is expected that in time intervals where the homogeneous concentration changes little, this Jacobian can be approximated by a constant value given by $\hat{\mathds{J}}=\frac{\partial \mathbf{f}} { \partial \mathbf{c}}(\mathbf{C}_0)$. This strategy simplifies the characterization of disturbances and allowed us to corroborate that their stability will depend on three factors: 1) the change in concentration with respect to the fixed point concentration; 2) the linear change in reaction rates induced by the dilution and 3) the direct effect of dilution as measured by local increase/decrease in volume. We show that an approximate criterion for the stability of perturbations is

\begin{equation}\label{eq:stableperturbations}
\lambda(t)= \mbox{Re}\{ \hat{\boldsymbol{\Lambda}}_i \}- \frac{1}{t}\log \frac{l(t)}{l(0)},
\end{equation}
to be negative for all the eigenvalues $\hat{\boldsymbol{\Lambda}}_i$ numbered by $i$ of the  Jacobian matrix evaluated at $\mathbf{C}_0$ given by $\hat{\mathds{J}}$.

\vspace{0.25cm}

In the next sections we will explain our methodology to find the Turing bifurcation from a potential function in successive stages. First, in Section \ref{sec:fixed}, we will present the idea of our potential function approximation to reproduce the well-known Turing conditions for a fixed domain obtained from the eigenvalue problem. Then in Section \ref{sec:constant}, we will derive the conditions for an RDD system where the steady-state concentration is approximated by $\mathbf{C}_0$, constant, as for exponential growth. In Section \ref{sec:numeric} we will test our hypotheses against extensive numerical solutions of the Brusselato RDD equations for this type og growing. Finally, in Section \ref{sec:discussion}, we provide our discussions and conclusions on the wavenumber selection problem.

\section{Turing conditions on a fixed domain in term of an energy function}\label{sec:fixed}

In this section we first summarize the Turing conditions of a reaction-diffusion system in a domain of fixed size $L$ for a two-component system from the traditional linear eigenvalue problem. Then, we will state the consequence of these conditions on a potential function that will allow us to state the general problem of finding the Turing bifurcation as the result of structural changes of this function.

The eq. \eqref{eq:perturbations} for fixed domain implies  $l'(t)=0$, $x=l(t) \xi=L \xi$ and  $\mathbf{C}_s \to \mathbf{c}_0$, and therefore, the perturbations obey 

\begin{equation}\label{eq:systemfixed}
 \frac{\partial \boldsymbol{\zeta}}{ \partial t} =   \mathds{D}  \frac{ \partial^2 \boldsymbol{\zeta}}{ \partial x^2}  +\mathds{J}\boldsymbol{\zeta}.
\end{equation}
Here $\mathds{J}$ is the Jacobian evaluated at the fixed point where $\mathbf{f}(\mathbf{c}_0)=\mathbf{0}$. It can be assumed periodic boundary conditions for the variable $ \boldsymbol{\zeta}$ between $0$ and $L$ and random small noise around $\mathbf{c}_0$ as initial conditions for the concentrations.

\subsection{Analysis in terms of eigenvalues}
The solution of \eqref{eq:systemfixed} consists of a linear combination of Fourier modes $\boldsymbol{\zeta}_k(x,t)= e^{i k x+\lambda t} \mathbf{v}^{(k)}$, where $k$ is the wavenumber, $\lambda_k$ an eigenvalue, and $\mathbf{v}^{(k)}$ the eigenvector of the matrix

\begin{equation}
\mathds{A}(k)=\mathds{J}-k^2 \mathds{D}.
\end{equation}
These eigenvalues are related to the dispersion relation

\begin{equation}\label{eq:dispersion}
\lambda_k^2 - \tau_{\mathds{A}}(k)\lambda_k+\Delta_{\mathds{A}}(k) =0,
\end{equation}
where $ \tau_{\mathds{A}}$ and $\Delta_{\mathds{A}}$ are the  trace and determinant of $\mathds{A}$, both depending explicitly on the wavenumber $k$.

In this case, the conditions for Turing pattern formation are: 

\begin{itemize}

\item  The system should be stable in absence of diffusion. This implies that the real part of the  eigenvalues of  $\mathds{A}$ for the mode with $k=0$ accomplishes $\mbox{Re}\{\lambda_0\}< 0$. From \eqref{eq:dispersion}, this requires

\begin{equation}\label{eq:fixed1}
\tau_\mathds{A}(0)<0 \mbox{ and }  \Delta_\mathds{A}(0)>0.
\end{equation}

\item  The system should be unstable when diffusion is turned on for at least some wavenumber  $k_m$. Since the previous condition on the trace implies that $\tau_\mathds{A}(k)<0$ for all $k$, the only way to unstabilize the system and obtain $\mbox{Re}\{\lambda(k_m)\}\geq 0$  is through

\begin{equation}\label{eq:fixed2}
\Delta_\mathds{A}(k_m)\leq 0  \mbox{ for some }  k_m>0.
\end{equation}

\end{itemize}

\vspace{0.5cm}

Let us assume that the related matrices on \eqref{eq:systemfixed} for a two component system are 

\begin{equation}\label{eq:matrices}
\mathds{J}=\left(\begin{array}{cc}
j_{11} & j_{12} \\ 
j_{21} & j_{22}
\end{array}\right)   \mbox{ and }  
\mathds{D}=\left(\begin{array}{cc}
d_{u} & 0\\ 
0 & d_v
\end{array}\right).
\end{equation}
In this terms, the first two conditions in \eqref{eq:fixed1} lead to $\tau_\mathds{J}(0)<0$ and  $\Delta_\mathds{J}>0$, whereas the second condition requires that the determinant given by $
\Delta_\mathds{A}(k,t)=k^4 \Delta_\mathds{D} - k^2 \sigma_{\mathds{D}\mathds{J}}+\Delta_\mathds{J}
$ has a minimum at $k_m$ where the function is negative. Here, $\sigma_{\mathds{DJ}}\equiv j_{11} d_v+j_{22} d_u$. The minimum of the determinant in the $k$ coordinate occurs at 

\begin{equation}
k_m=\sqrt{\frac{\sigma_{\mathds{D}\mathds{J}}}{2 \Delta_{\mathds{D}}}}.
\end{equation}
  where it has the value $\Delta_\mathds{A}(k_m,t)=\Delta_{\mathds{J}} -\frac{\sigma_{\mathds{D}\mathds{J}}^2}{2 \Delta_{\mathds{D}}}$. Therefore,  the conditions on\eqref{eq:fixed2} in terms of the original matrix are 
 
\begin{equation}\label{eq:fixedunstable}
\sigma_{\mathds{DJ}}>0 \mbox{ and }  \sigma_{\mathds{DJ}}^2- 4 \Delta_{\mathds{D}}\Delta_\mathds{J} >0.
\end{equation}

The width of wavenumbers where $\lambda_k$ has positive real part, and where patterns can occur, is given by 

\begin{equation}
|k^2-k_m^2|\leq \sqrt{k_m^4 - \frac{\Delta_{\mathds{J}}}{\Delta_{\mathds{D}}}}.
\end{equation}
If some of the two diffusion coefficients on $\Delta_{\mathds{D}}$ is used as a bifurcation parameter, then Turing bifurcation occurs when $\Delta_{\mathds{D}}^b=\sigma_{\mathds{DJ}}^2/4 \Delta_{\mathds{J}}$ at the wavenumber $k_b=\sqrt{2 \Delta_{\mathds{J}}/\sigma_{\mathds{DJ}} }$. A detailed computation of these conditions is standard and given for example at Ref. \cite{murray2001mathematical}.

\subsection{Analysis in terms of potential functions}

In component form, the Fourier mode of the perturbation on \eqref{eq:systemfixed} $\boldsymbol{\zeta}_k=(u_k,v_k)^T$  becomes

\begin{align*}
u'_k(t)&=-k^2 d_u u_k + j_{11} u_k + j_{12} v_k,\\
v'_k(t)&=-k^2 d_v v_k + j_{21} u_k + j_{22} v_k.\\
\end{align*}
After some calculations, this system can be written as a second order differential equation

\begin{equation}\label{eq:uk1}
u_k''(t)-\tau_\mathds{A}(k) u'_k+  \Delta_\mathds{A}(k) u_k=0,
\end{equation}
and exactly the same equation holds for $v_k$. The characteristic equations  would lead  to the dispersion relation in \eqref{eq:dispersion}, and therefore to the same conditions on  \eqref{eq:fixed1} and \eqref{eq:fixed2}.

Now let us understand the Turing conditions differently. If we multiply \eqref{eq:uk1} by $u'_k$, the equation for the $k-$esim Fourier is

\begin{equation}
\frac{d V_k}{dt}=\tau_\mathds{A}(k)  u'^2.
\end{equation}
where the function $V_k$ is given by

\begin{equation}
 V_k(u,u') \equiv \frac{u'^2}{2} + \Delta_\mathds{A}(k) \frac{u^2}{2}.
\end{equation}

When the  conditions \eqref{eq:fixed1}   for stability in absence of diffusion apply ($k=0$), this function for the mode $k=0$, is an elliptic paraboloid with minimum at the origin, always positive and accomplishing $\frac{d}{dt}V_0\leq 0$. Therefore $V_0$ accomplish the conditions for a Lyapunov function  and the origin is asymptotically stable in absence of diffusion.  

Now,  when diffusion is turned on and the conditions in \eqref{eq:fixed2} holds, then $\frac{d}{dt}V_{k_m} \leq 0$, but now $V_{k_m}$ is a saddle. Therefore, if $\mathbf{u}_k=(u_k,u_k')$, as the relation 

\begin{equation}
\frac{dV_k}{dt}=\frac{dV_k}{d\mathbf{u}_k}\cdot \frac{d\mathbf{u}_k}{dt}
\end{equation} 
holds for any wavenumber $k$, then the condition $\frac{d V_{k_m}}{dt} \leq 0$ implies that there is at least one trajectory that diverges as the time grows, making the origin $\mathbf{u}_{k_m}=\mathbf{0}$ necessarily unstable for this wavenumber. 

In conclusion, the well-known conditions for the occurrence of a Turing pattern in a fixed domain given by \eqref{eq:fixed1} and \eqref{eq:fixed2} imply that, as increasing $k$, the family of functions $V_k$ deforms from an elliptical paraboloid (with $k=0$) to a saddle (for wavenumbers around $k_m$) along which the trajectories related to that Fourier mode diverge; passing this interval, the functions are again elliptic paraboloids where the trajectories associated to the wavenumbers are stable. In all cases, the dynamics of the system is such that the the value of $V(t)$ always decrease with time, and therefore, can be interpreted as a kind of potential function for each Fourier mode numbered by  $k$. This form of potential is valid only in the neighborhood of the origin where the linear approximation of each Fourier mode holds, and the nonlinear terms are expected to produce saturation containing the growth of the amplitudes of the unstable modes, as occurs in a fixed domain.It will result physical insightful how these particular function can be associated to a single  potential function of the entire reaction-diffusion process that takes into account all the modes \cite{ledesma2022energy}. 

The hypothesis of the work for finding the Turing conditions for a growing domain is the following: there are a set of conditions (similar to \eqref{eq:fixed1} and \eqref{eq:fixed2}) for the emergence of unstable modes  that we do not know, but whose effect on a potential functions is the same: to deform from an elliptic paraboloid to a saddle and again to a paraboloid when increasing  the wavenumber; therefore, some conditions  for Turing formation can be guessed from the behaviour of the potential functions. In the following  section we will provide a methodology to obtain the equivalent of these two conditions for a system that grows with time. Then we will test by numerical simulations if the proposed conditions predict Turing patterns.

 \section{Turing conditions for growing domain with constant homogeneous concentration}\label{sec:constant}

Let us consider first the case where $\mathbf{c}_s\approx \mathbf{C}_0$ and therefore $ \frac{\partial \mathbf{f}(\mathbf{c}_s)}{\partial  \mathbf{c}}\approx  \frac{\partial \mathbf{f}(\mathbf{C}_0)}{\partial  \mathbf{c}} = \hat{\mathds{J}}$.  In this case, after taking the Fourier transform in the  computational domain, for each wavenumber $\kappa$, eq. \eqref{eq:system} becomes

\begin{equation}
 \frac{\partial \boldsymbol{\zeta}_\kappa}{ \partial t}  = \left[ \hat{ \mathds{J}}-\left(\frac{\kappa}{l(t)}\right)^2\mathds{D} -\frac{\dot{l}(t)}{l(t)} \mathds{I} \right] \boldsymbol{\zeta}_\kappa,
\end{equation}
In order to simplify notation, we can use $k(t) \equiv \kappa/l(t)$ as the wavenumber in the actual domain and $g(t)\equiv \dot{l}(t)/l(t)$. The matrix in parenthesis is therefore:

\begin{equation}\label{eq:caligraphcA}
A(\kappa,t)= \hat{ \mathds{J}}-k^2(t) \mathds{D} -g(t) \mathds{I}. 
\end{equation}

\subsection{Potential function}
 If $\boldsymbol{\zeta}_\kappa=(u_\kappa,v_\kappa)$, this system in component form is 

\begin{align*}
u'_\kappa(t)&=-\left(\frac{\kappa}{l(t)}\right)^2 d_1 u_\kappa + j_{11} u + j_{12} v_\kappa-g(t) u_k,\\
v'_\kappa(t)&=-\left(\frac{\kappa}{l(t)}\right)^2  d_2 v_\kappa + j_{21} u_\kappa + j_{22} v_\kappa-g(t) v_k.\\
\end{align*}
The second order equation for $u$ is 

\begin{equation}\label{eq:uequation}
u''_\kappa(t)-\tau_A(\kappa,t) u'_\kappa+ \left[ \Delta_A(\kappa,t)  +2 d_u k(t) k'(t) +g'(t)  \right] u_\kappa=0,
\end{equation}
and one similar for $v_k$ changing $d_u \to d_v$. Multiplying by $u'_{\kappa}$ and rearranging, we get 

\begin{equation}
\frac{d}{dt} \left\{ \frac{u_\kappa'^2}{2} + \left[ \Delta_A(\kappa,t) +2 d_u k(t) k'(t)  + g'(t)    \right] \frac{u_\kappa^2}{2}  \right\}=\tau_A(\kappa,t)  u_\kappa'^2+ \frac{u^2_\kappa}{2} \frac{d}{dt} \left[  \Delta_A(\kappa,t) +2 d_u k(t) k'(t) +g'(t)    \right] .
\end{equation}
From this equation is clear that the potential function now is 

\begin{equation}
V_{\kappa}=\frac{u_\kappa'^2}{2} + \left[ \Delta_A(\kappa,t) +2 d_u k(t) k'(t)  + g'(t)\right] \frac{u_\kappa^2}{2}\end{equation},
and its time rate is 

\begin{equation}\label{eq:dotV}
\dot{V}_{\kappa}=\tau_A(\kappa,t)  u_\kappa'^2+ \frac{u^2_\kappa}{2} \frac{d}{dt} \left[  \Delta_A(\kappa,t) +2 d_u k(t) k'(t) +g'(t)    \right].
\end{equation}

The conditions for stability in absence of diffusion ($\kappa=0$) require $V_0\geq 0$ and $\dot{V}_0 \leq 0$.  This implies that 

\begin{equation}\label{eq:constant1}
 \Delta_A (0,t)+g'(t) \geq 0 \mbox{ , } \tau_A(0,t) \leq 0 \mbox{ and }  \frac{d}{dt} \left[ \Delta_A(0,t) +g'(t) \right]  \leq 0.
\end{equation}
These conditions guarantee that $V_0$ is an elliptical paraboloid centered at the origin, towards which all trajectories are directed. However, care must be taken since, unlike what happens in the fixed domain, the paraboloid defined by  

\begin{equation}
V_0=\frac{u'^2}{2} +[ \Delta_\mathcal{A}(0,t) +g(t)]  \frac{u^2}{2},
\end{equation}
is changing its width with  time, and therefore \eqref{eq:constant1} could not necessarily guarantee that all trajectories reach the origin. To do this, one would have to add a condition that the rate at which the width of the paraboloid increases is slower than the rate at which the mode decays to zero.

Instability with diffusion requires  that for some $\kappa_m$, $V_{m} \geq 0$ and that $V_m$ is a saddle, which requires 

\begin{equation}\label{eq:constant2}
 \Delta_\mathcal{A}(\kappa_m,t) +2 d_u k_m(t) k_m'(t)  +g'(t)    <0. 
\end{equation}
The condition for $V_m$ to be a potential is that all the trajectories descend,  $\dot{V}_{m} \leq 0$, this requires the extra conditions 

\begin{equation}\label{eq:constant3}
\tau_A(\kappa_m,t)  <0 \mbox{ and } \frac{d}{dt} \left[  \Delta_A(\kappa_m,t) +2 d_u k_m(t) k_m'(t) +g'(t)  \right]  <0.
\end{equation}
This second set of conditions for unstability is also interesting because implies that the unstabilization of a mode can be due to a change in the structure of the potential function $V$ from a paraboloid to a saddle, as we have supposed on \eqref{eq:constant2},  but also due to a change in the sign of $\dot{V}$ in \eqref{eq:dotV}. We will follow the first line since this replicates what occur for fixed domain, and left the second possibility  for future studio.  

In conclusion, due to these details, the set of conditions \eqref{eq:constant1} and \eqref{eq:constant2} can be seen for now as an approximate set of Turing conditions, neither necessary nor sufficient, but as we shall see, gives a very good idea of the Turing region in the parameter space and will be proved numerically in the following sections. Now we focus on the relation between the conditions on  \eqref{eq:constant3} and the relation with the original RDD system.

\subsection{Turing conditions in terms of the original matrix}
From \eqref{eq:caligraphcA}, trace and determinant of $A$ are respectively
\begin{equation}
\tau_A(\kappa,t)= \tau_\mathds{\hat{J}}- k^2(t) \tau_\mathds{D}-2g(t),
\end{equation}

\begin{equation}
\Delta_\mathcal{A}(\kappa,t)=  \Delta_\mathds{D} k^4(t) -k^2(t) \sigma_{\mathds{D}\mathds{\hat{J}}}+\Delta_\mathds{\hat{J}} +[k^2(t) \tau_{\mathds{D}}-\tau_{\mathds{\hat{J}}}] g(t)+g^2(t).
\end{equation}

In this terms, the conditions for stability with $\kappa=0$ in \eqref{eq:constant1} requires
\begin{eqnarray}
\Delta_\mathds{\hat{J}} -\tau_{\mathds{\hat{J}}} g(t)+ g^2(t)+ g'(t) && \geq 0, \nonumber\\
    \tau_\mathds{\hat{J}}-2g(t)&& \leq 0, \\
 -g'(t) [\tau_\mathds{\hat{J}}-2 g(t)] + g''(t) && \leq 0.  \nonumber
\end{eqnarray}

The condition \eqref{eq:constant2}   for instability require that  the function 
\begin{equation}\label{eq:hu}
H_u(k)= k^4(t) \Delta_\mathds{D} + [ \tau_{\mathds{D}} g(t)- \sigma_{\mathds{D\hat{J}}}] k^2(t) +2 d_u k(t) k'(t) 
+ \Delta_{\mathds{\hat{J}}}-\tau_{\mathds{\hat{J}}} g(t)+ g^2(t) +g'(t),
\end{equation}
to be negative. To find the minimum wavenumber, we return this expression to the wavenumber $\kappa$, derivate with respect to this variable, and return the result to the actual wavenumber. The minimum of $H_u$ occurs at 

\begin{equation}\label{eq:wavenumberc}
k_u(t)=\sqrt{\frac{(2 d_u-\tau_{\mathds{D}}) g(t) + \sigma_{\mathds{D\hat{J}}}}{2 \Delta_{\mathds{D}}}}.
\end{equation}
Therefore, the existence of the minimum of $H_u$  requires
\begin{equation}
(2 d_u-\tau_{\mathds{D}}) g(t) + \sigma_{\mathds{D\hat{J}}}>0.
\end{equation}
Since the value of the function $H_u$ must be negative at $k_u$, 

\begin{equation} 
\sigma_{\mathds{D\hat{J}}}^2- 4 \Delta_{\mathds{D}}\Delta_\mathds{\hat{J}} +2g(t)[ (2 d_u-\tau_{D})\sigma_{\mathds{D\hat{J}}} +2  \tau_{\mathds{D}}\Delta_\mathds{\hat{J}}]+g^2(t)[4 d_u^2-4 d_u \tau_{\mathds{D}}-4 \Delta_{\mathds{D}}+\tau_{\mathds{D}}^2] -4\Delta_{\mathds{D}} g'(t)
> 0.
\end{equation}
The possible wavenumbers where unstability occurs are those with
\begin{equation}\label{eq:widthc}
|k^2-k_u^2| \leq \sqrt{ k_u^4 -\frac{1}{\Delta_\mathds{D}} [\Delta_{\mathds{J}}+g'(t)-\tau_\mathds{J} g(t)+g(t)^2] }, 
\end{equation}
 and the bifurcation value is

\begin{equation}\label{eq:bifurcationu}
\Delta_\mathds{D}^{u,b}=\frac{[(2 d_u-\tau_\mathds{D}) g(t)+\sigma_\mathds{DJ}]^2}{4 \left[\Delta_\mathds{J}+g'(t)- \tau_\mathds{J} g(t)+g(t)^2\right]}
\end{equation}

Two aspects are worth mentioning in this deduction. The first is related to the concentration dependence $u $ through the coefficient $d_u $ in the equations \eqref{eq:hu}-\eqref{eq:bifurcationu}. Since the concentration $v$ obeys an equation similar to \eqref{eq:uequation} but with $d_v$, then the same reasoning applies substituting $d_v$. If one is only interested in the conditions for the occurrence of patterns, then it is enough to ask, for $I4$, for example, that $\min_i \{(2 d_i-\tau_{\mathds{D}}) g(t) + \sigma_{\mathds{D\hat{J}}} \}>0$, and something similar for the condition $I5$. Once the diffusion coefficient (where the condition of minimum applies) is selected, it is possible to predict if the wave number is $k_u$ or $k_v$. This will be illustrated by the example of the Brusselator.

In Table \ref{tab:conditions} we summarize the conditions for Turing pattern emergency in fixed and growing domains. Notice that the right column can recover the well conditions for a fixed domain in the left when the growth rate is zero $g(t)\to 0$. For simplicity we have added the labels $S$, $I$ and $D$, pointing to stability, unstability and domain conditions, respectively.

\begin{table*}[h!]
\begin{tabular}{|C{1cm}|C{7cm}|C{9cm}|}
\hline 
\# & Fixed Domain & Growing Domain \\ 
\hline 
\hline
S1) & $\Delta_\mathds{J}>0$  & $\Delta_\mathds{\hat{J}} -\tau_{\mathds{\hat{J}}} g(t)+ g^2(t)+g'(t)  >0$ \\ 
\hline 
S2)  & $\tau_\mathds{J}<0$ & $\tau_\mathds{\hat{J}}-2g(t)<0$ \\ 
\hline 
D3) & - - -  & $g'(t) [2g(t)-\tau_\mathds{\hat{J}}]+g''(t) \leq 0$  \\ 
\hline
\hline 
I4) & $ \sigma_{\mathds{DJ}}>0$ & $(2 d_i-\tau_{\mathds{D}}) g(t) + \sigma_{\mathds{D\hat{J}}}>0$  \\ 
\hline 
I5) & $\sigma_{\mathds{DJ}}^2- 4 \Delta_{\mathds{D}}\Delta_\mathds{J} \geq 0$ 
& $\sigma_{\mathds{D\hat{J}}}^2- 4 \Delta_{\mathds{D}}\Delta_\mathds{\hat{J}} +2g(t)[ (2 d_j-\tau_{D})\sigma_{\mathds{D\hat{J}}} +2  \tau_{\mathds{\hat{J}}}\Delta_\mathds{D}]$  \\ 
 & & $+g^2(t)[4 d_i^2-4 d_i \tau_{\mathds{D}}-4 \Delta_{\mathds{D}}+\tau_{\mathds{D}}^2] -4\Delta_{\mathds{D}} g'(t) \geq 0$  \\ 
 \hline
 \hline
 $k_m$ &  $\sqrt{ \frac{ \sigma_{\mathds{DJ}}}{4 \Delta_\mathds{D}} }$  & $\min_i  \left \{  \sqrt{\frac{(2 d_i-\tau_{\mathds{D}}) g(t) + \sigma_{\mathds{D\hat{J}}}}{2 \Delta_{\mathds{D}}}} \right\} $   \\
\hline
$\delta k^2$  & $\sqrt{ k_m ^4 - \frac{\Delta_{\mathds{J}}}{\Delta_\mathds{D}}}$  &   
$\sqrt{ k_m^4 -\frac{ [\Delta_{\mathds{\hat{J}}}+g'(t)-\tau_\mathds{\hat{J}} g(t)+g(t)^2]}{\Delta_\mathds{D}} }$ \\
\hline
$\Delta_{\mathds{D}}^b$ &
$\frac{4 k_m^4 \Delta^2_{\mathds{D}}}{ \Delta_{\mathds{J}}}$
& $\frac{4 k_m^4 \Delta^2_{\mathds{D}}}{ \left(\Delta_{\mathds{\hat{J}}}+g'(t)-\tau_{\mathds{\hat{J}}} g(t)+2 g(t)^2\right)}$ \\
\hline
\end{tabular} 
 \caption{Turing conditions for a two component system with iotropic growth and constant homogeneous state. Center column summarize the well known Turing conditions for fixed domain in Section \ref{sec:fixed}. Right column presents the equivalent conditions for growing domain. $\tau$ and $\Delta$ refer to trace and determinant of the matrix in the subscript, as it does $\sigma$ , which can be the Jacobian $\mathds{J}$ at the fixed point, the diagonal diffusion $\mathds{D}$, or the Jacobian at the representative concentration $\mathds{\hat{J}}$, all constant matrices. }\label{tab:conditions}
 \end{table*}
 
The second aspect is much more delicate and has to do with the explicit appearance of time in the equations. In the following Part of the work, we will reveal through  numerical examples some observations made in this regard and its role as a parameter of the Turing bifurcation. However, as a start, the exponential growth presents a simple situation since $g(t)$ is constant, therefore many terms in the Turing conditions become zero, plus the Turing conditions can be studied independent of time.  Therefore, it will be our first case study.

\section{The exponential case}\label{sec:numeric}
Consider the case where growth/shrinkage is exponential $l(t)=l(0) e^{rt}$, where the growth rate is $g(t)=r$, constant, and all derivatives of $g$ , they become zero. As we have explained in Section \ref{sec:antecedents}, the homogeneous state of exponential growth ($r>0$) and shrinkage ($r<0$) quickly tends to a constant value. We showed in Part 1 of this paper that if $\mathbf{c}_0$ is the fixed point of the reactive system where $f(\mathbf{c}_0)=\mathbf{0}$, then an approximation for the value representative in \eqref{eq:representative}, if $|r|$ is relatively small,  is

\begin{equation}\label{eq:representativeexpo}
\mathbf{C}_0=[\mathds{I} +r(\mathds{J}-r\mathds{I}) ] \mathbf{c}_0.
\end{equation} 

The importance of the non-linear term of $\mathbf{f}(\mathbf{c})$ in finding the homogeneous state of \eqref{eq:homogeneous} lies in the fact that the dilution term can induce new equilibrium points for large values of $|r|$. To see this, notice that if $g(t)=r$ in eq. \eqref{eq:representative}, the homogeneous state can be studied as

\begin{equation}\label{eq:homogeneous2}
 \frac{\partial   \mathbf{c}_s}{ \partial t} =\hat{\mathbf{f}}(\mathbf{c}_s),
\end{equation} 

\noindent where $\hat{\mathbf{f}}(\mathbf{c}_s)\equiv \mathbf{f}(\mathbf{c}_s)- r \mathbf{c}_s$ and therefore, it is enough to study the fixed points of the modified reaction $\hat{\mathbf{f}}(\mathbf{c}_s)$. Let us call this points  $\mathbf{\hat{c}}_0$, where $\hat{\mathbf{f}}(\mathbf{\hat{c}}_0)=\mathbf{0}$ \cite{gjorgjieva2007turing}. In other words, eq. \eqref{eq:homogeneous2} shows that the dilution term for exponential growth has the same effect that adding two linear reactions of the same rate constant $r$ for each concentration. As we will illustrate for our study cases, this can change the number of fixed points and also its stability.

\subsection{Brusselator example}

\subsubsection{Fixed point concentration}
The Brusselator is given by \eqref{eq:bruselas1}.  The fixed point of the  reaction  itself is at $\mathbf{c}_0=(A,B/A)^T$, and  the Jacobian and diffusion matrix of the fixed-domain problem in \eqref{eq:matrices} (see Ref. \cite{ledesma2020eckhaus}) are 

\begin{equation}
\mathds{J}=\left(\begin{array}{cc}
-1+B & A^2 \\ 
-B & -A^2
\end{array}\right)   \mbox{ and }  
\mathds{D}=\left(\begin{array}{cc}
\sigma & 0\\ 
0 & 1
\end{array}\right).
\end{equation}

\begin{figure*}[hbtp]
\centering
\includegraphics[width=1 \textwidth ]{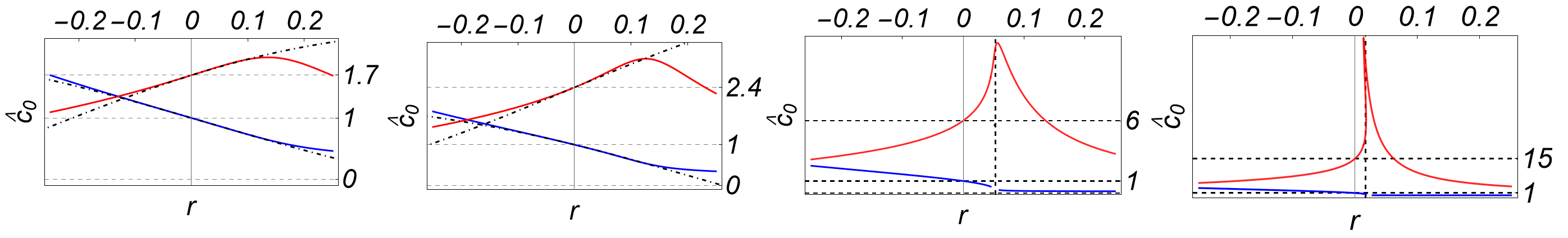}
\caption{Real valued fixed point concentrations. In solid color we obtain that obtained solving numerically $\hat{\mathbf{f}}(\mathbf{\hat{c}}_0)=\mathbf{0}$. In dot-dashed lines we plot the approximation for \eqref{eq:aproximaC0}. The values of the Brusselator in \eqref{eq:bruselas2} used where $A=1$ and $B=\{1.7,2.4,6,15\}$, respectively.  \label{fig:fixedpoints}}
\end{figure*}

\noindent When the domain grows and dilution is included,  from \eqref{eq:homogeneous2}, the equation describing the homogeneous state is

\begin{align}\label{eq:bruselas2}
\frac{\partial c_u}{\partial t}&=A-Bc_u-c_u +c_u^2 c_v -r c_u,\\
\frac{\partial c_v}{\partial t}&= Bc_u -c_u^2 c_v-r c_v.
\end{align}
Given the non-linearity of the equations, the system can have from one to five equilibrium points depending on the values of the parameters. In Fig. \ref{fig:fixedpoints}, we illustrate the real-valued fixed point $\mathbf{\hat{c}}_0$ for some combinations of parameters.  

We will study the situation when the fixed point changes slowly as a function of the growth parameter $r$. In particular, for values of $|r|$ close to zero, the linear approximation of the representative concentration at \eqref{eq:representativeexpo} is:

{\small
 \begin{equation}\label{eq:aproximaC0} 
  \mathbf{C}_0=\left(\frac{A \left(A^2-2 B r+r\right)}{A^2 (r+1)+r (-B+r+1)}  ,  \frac{A B (2 r+1)}{A^2 (r+1)+r (-B+r+1)}\right)^T  
 \end{equation}
}
\noindent This value is plotted only for the first two columns in Fig. \ref{fig:fixedpoints} (in black lines) and compared with the numerical results giving good results between $r=-0.15$ and $0.15$. Notice that by increasing the value of $B$, the system goes from a single fixed point ($B$ up to 6) to three ($B=15$). We will focus only on parameter values that result in a single fixed-point real-valued concentration. Using numerical approximations up to the second order in $r$, near $r\approx 0$, the approximation of the fixed concentration of the system at  \eqref{eq:bruselas2} (in blue and red in Fig.  \ref{fig:fixedpoints}) is

{\small
 \begin{equation}\label{eq:aproximavc0hat} 
  \mathbf{\hat{c}}_0\approx \left(A+\frac{r \left(-A^2-B\right)}{A}+\frac{r^2 \left(A^4-B^2+B\right)}{A^3}  ,\frac{ B}{A} +\frac{B (-1 + A^2 + B) r}{A^3} + \frac{
 B (1 - 4 B + 2 B^2 + A^2 (-3 + 2 B)) r^2}{A^5}\right)^T  
 \end{equation}
}

\noindent which matches \eqref{eq:aproximaC0} to first order but approximates better for values larger of $|r|$. In what follows, we use the fixed point obtained numerically directly from solving $\hat{\mathbf{f}}(  \mathbf{\hat{c}}_0)=\mathbf{0}$ for our graphic results on Turing patterns and keep the approximation in \eqref{eq:aproximavc0hat} for analytic expressions.

\subsubsection{Turing space}
\begin{figure*}[hbtp]
\centering
\includegraphics[width=0.71 \textwidth]{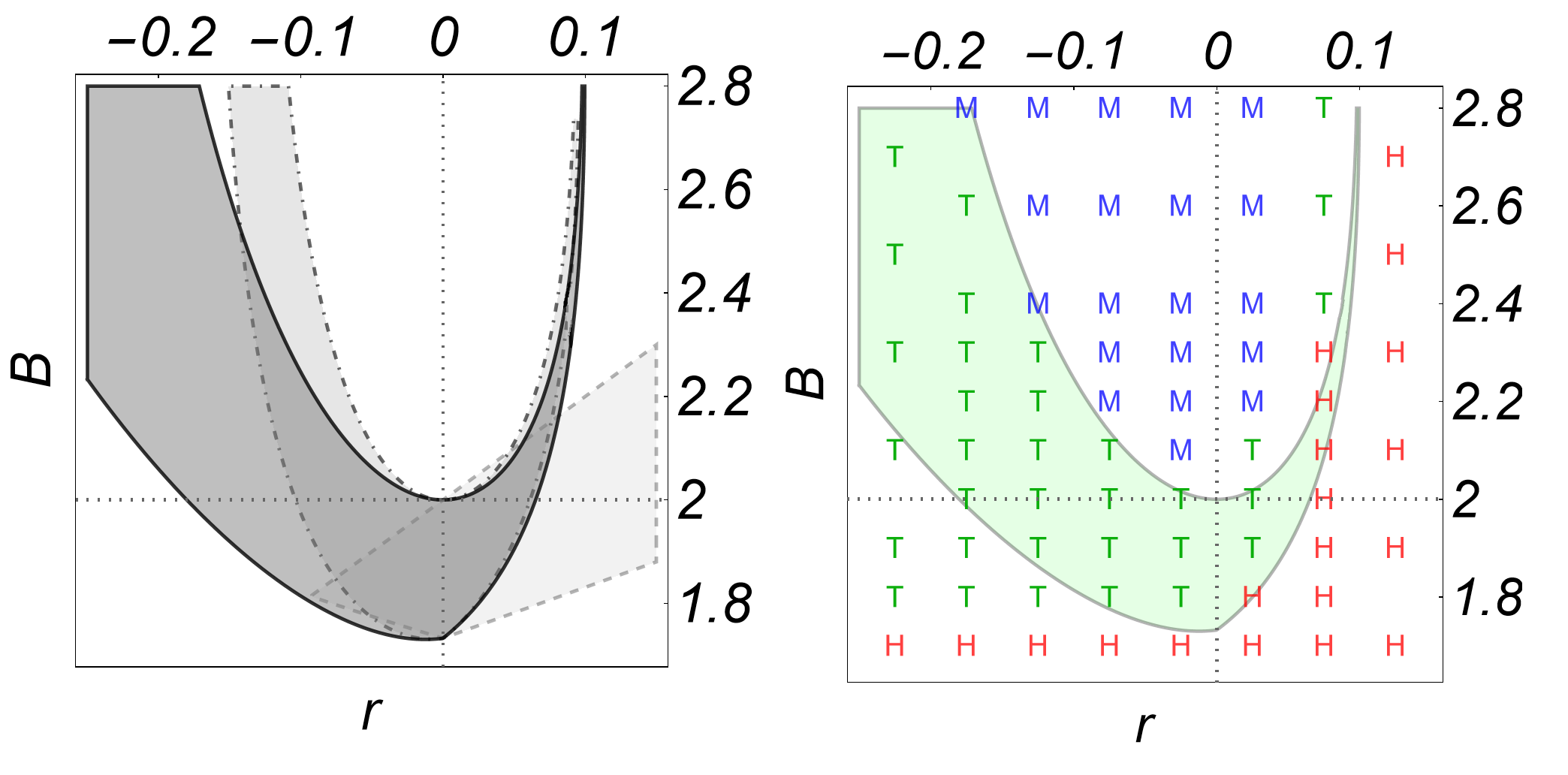}
\caption{Turing space of the Brusselator for exponential growth. a) Prediction of our model using as fixed point concentration: $\mathbf{\hat{c}}_0$ (solid boundary), $\mathbf{C}_0$ (dotdashed boundary) and $\mathbf{c}_0$ (dashed boundary). b) Homogeneous (H), Turing (T) and Mixed mode solutions (MM) as derived from numerical simulations.  \label{fig:turingspace}}
\end{figure*}

Using the approximation to the fixed-point at \eqref{eq:aproximavc0hat}, the Turing conditions  in Table \ref{tab:conditions} for the Brusselator and valid for low values of $|r|$ are given in Table \ref{tab:bruselas}. The full expression for these conditions using, in turn, the full expressions for $\mathbf{\hat{c}}_0$ is too big to write on one page, but they are the ones considered to build Turing space in Figure \ref{fig:turingspace}.a. Setting $A=1$ and $\sigma=0.1$, we consider the variation of the parameters $(r,B)$ to study the effect of growth and distance to the bifurcation, respectively. Furthermore, to distinguish the different orders of approximations, in Fig. \ref{fig:turingspace}.a, we also construct the Turing space using the complete expressions for $\mathbf{\hat{c}}_0$ (solid boundary region), using the approximation $\mathbf{C}_0$ (dotdashed boundary) and the fixed point of the pure reaction $\mathbf{c}_0$ (dashed boundary).

\begin{table*}[h!]
\begin{tabular}{|C{1cm}|C{16cm}|}
\hline 
\# & Approximate expressions for the Turing conditions in Table \ref{tab:conditions}  \\ 
\hline 
\hline
S1) &  $A^2+r \left(-A^2-3 B+1\right)+r^2 \left(-\frac{(B-4) B}{A^2}+A^2+1\right)>0$ \\ 
\hline 
S2)  & $(B-1-A^2) +\frac{2 \left(A^2-1\right)  \left(A^2+B\right)r}{A^2}+r^2 \left(\frac{2 B (1-2 B)}{A^4}+\frac{(B-6) B}{A^2}-3 A^2-2 B\right)<0$ \\ 
\hline 
I4a) & $(B-1-A^2 \sigma)+r \left(2 \sigma  \left(A^2+B\right)-\frac{2 B}{A^2}+\sigma -1\right)+\frac{r^2 \left(-3 A^6 \sigma +B^2 \left(A^2 \sigma -4\right)-2 B \left(A^4 \sigma +A^2 (\sigma +2)-1\right)\right)}{A^4} \geq 0$  \\
\hline 
I4b) & $(B-1-A^2 \sigma)+r \left(2 \sigma  \left(A^2+B\right)-\frac{2 B}{A^2}-\sigma +1\right)+\frac{r^2 \left(-3 A^6 \sigma +B^2 \left(A^2 \sigma -4\right)-2 B \left(A^4 \sigma +A^2 (\sigma +2)-1\right)\right)}{A^4} \geq 0$  \\
\hline
I5a) & $\left[ \left(A^2 \sigma -B+1\right)^2-4 A^2 \sigma \right]-\frac{2 r \left(2 A^6 \sigma ^2+A^4 (2 B+1) (\sigma -1) \sigma +A^2 \left(-2 B^2 \sigma -7 B \sigma +B+3 \sigma -1\right)+2 (B-1) B\right)}{A^2} \geq 0$\\
\hline
I5b) & $\left[ \left(A^2 \sigma -B+1\right)^2-4 A^2 \sigma \right]-\frac{2 r \left(2 A^6 \sigma ^2+A^4 (2 B-1) (\sigma -1) \sigma +A^2 \left(-2 B^2 \sigma -B (5 \sigma +1)+\sigma +1\right)+2 (B-1) B\right)}{A^2} \geq 0$\\
\hline
\end{tabular} 
 \caption{Approximate Turing conditions for the Brusselator for exponential growth, valid near $|r| \approx 0$.  For the exponential growth, the condition D3 in Table \ref{tab:conditions} fulfils trivially since derivatives of $g(t)$ are null.} \label{tab:bruselas}
 \end{table*}

To test these predictions, in Fig. \ref{fig:turingspace}.b we present the results of the numerical simulations performed in Comsol Multiphysics of the RDD system at \eqref{eq:system} for the Brusselator system at \eqref{eq:bruselas1} for exponential growth. The simulations are performed in a fixed computational domain with 100 equidistant vertices with a simulation time $t_{max}$ calculated as the time it takes for the system to grow/shrink ten times the original size, depending on whether it is growing/shrinking, respectively, \emph{i.e.}, $t_{max}=|(1/r)\,\log(10)|$. The initial domain size is calculated using as reference the bifurcation wavenumber in a fixed domain ($r=0$), $k_c=\sqrt{ A/ \sqrt{\sigma }}$, and the expression $l (0)=2n\pi/k_c$. with $n$ equal to 3 or 19 for $r>0$ and $r<0$, respectively. We have used periodic boundary conditions and random disturbances of 10\% of the value of the initial concentration of $\mathbf{\hat{c}}_0$.

In Figure \ref{fig:turingspace}, the symbols H, T, and M represent homogeneous, Turing, and mixed-mode solutions, respectively. Homogeneous refers to the solution in which the initial spatial disturbance disappears and the system returns to fixed-point concentration over the entire domain. Turing solutions are those in which the system has a non-zero amplitude wavenumber, and spatial oscillations occur around the constant fixed-point concentrations. Mixed-mode patterns are solutions that are not properly a Turing pattern but have the same quality of being solutions with a very distinctive wavenumber with non-zero amplitude; the difference with Turing patterns is that the spatial pattern oscillates around not a fixed concentration, but around a limit cycle derived from proximity of the Hopf bifurcation. These solutions have already been found for fixed domains in  \cite{challenger2015turing,ledesma2019primary,aragon2012nonlinear}, and reported for growing domains in \cite{van2021turing}.

As can be seen in Fig. \ref{fig:turingspace}.b, the region predicted by our scheme summarized in Table \ref{tab:conditions} provides a good approximation of the Turing space found in the simulations. This region presents two asymmetries: the first concerns the Turing bifurcation line, which is constituted by the two parabolas that intersect at $r=0$, and it is different on each side due to the duplication of conditions for the different coefficients of diffusion; and the second concerning the Hopf bifurcation line, which is the upper parabola slightly tilted to the left. For the Brusselator, this combination makes the Turing space wider for shrinking than for growing, as predicted by our scheme and confirmed by numerical simulations.

Regarding the differences between our prediction and the numerical simulations, especially regarding the borders of the Turing region with homogeneous solutions, it should be noted that these are minimal for low values of $|r| \lesssim 0.1$, and arise for high values of $|r|$ and that could be due to the criteria used to distinguish between both solutions. To make the distinction, we proposed that a solution is homogeneous if its time-averaged amplitude is less than $0.01$. This means that for more negative values of $r$ (left side), the T symbols at the bottom may be overestimated because the simulation time is too short; the domain is reduced very quickly and our criteria may not be sufficient to ensure that the initial disturbance has been homogenized or not. As for the horn-shaped region on the right-hand side (more positive values of $r$), the transition between the Homogeneous, Turing, and Mixed mode solutions occurs very quickly as the parameter $B$ increases and also in a less differentiated way, so it is difficult to set strict limits with the type of simulation considered here. Therefore, the simulations carried out, rather than giving a strict characterization, allow us to confirm the essential characteristics of Turing space in terms of its location,  growth/shrink asymmetry, and range of occurrence between both bifurcations.

\begin{figure*}[hbtp]
\centering
\includegraphics[width=0.75 \textwidth ]{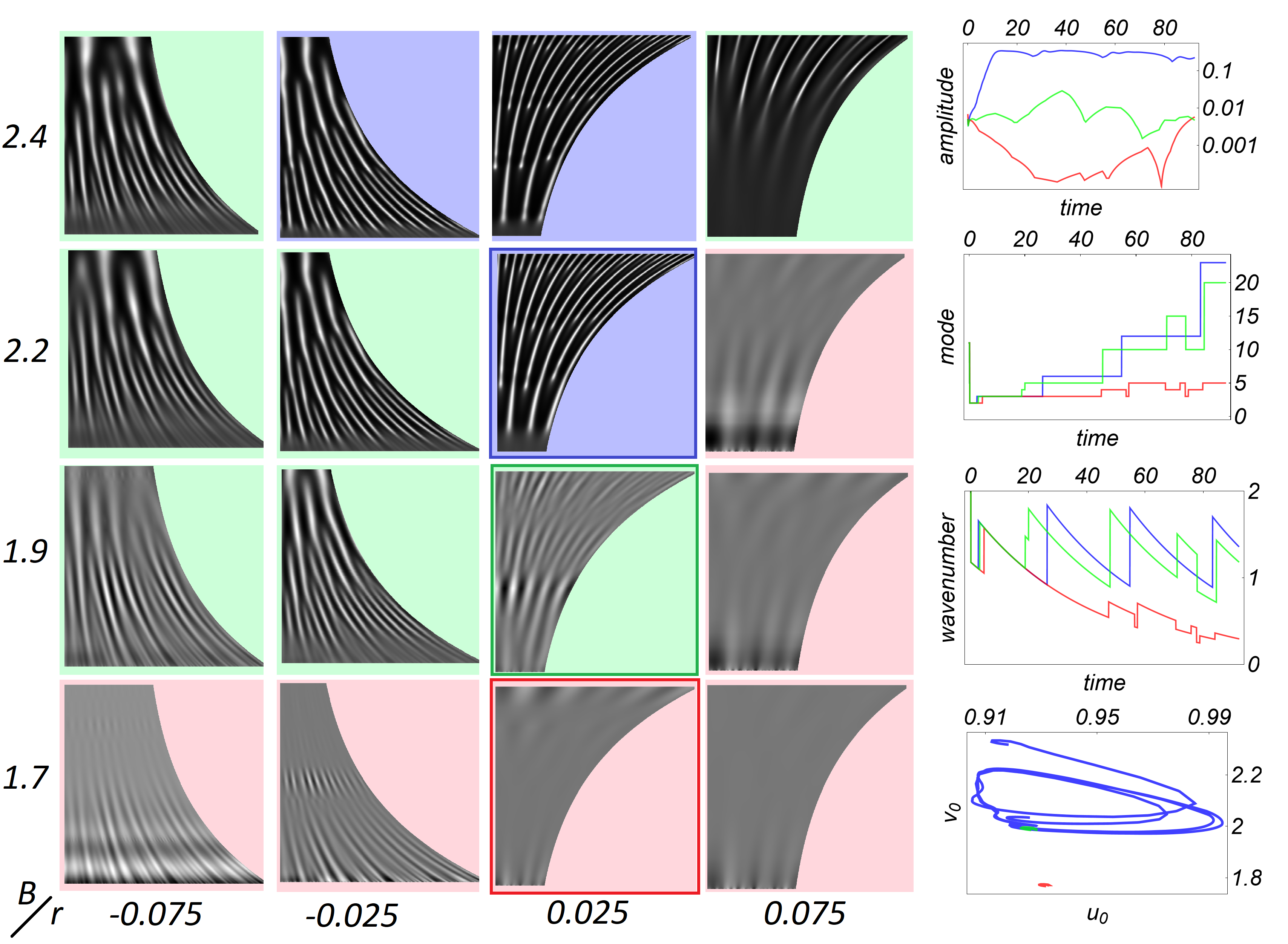}
\caption{ Different types of solution. a) Spatio-temporal maps (not to scale) of the solutions Homogeneous (red), Turing (green) and Mixed-mode (blue) solutions for different values of $(r,B)$. Example of amplitude, preponderant mode, actual wavenumber and phase space of the zero order Fourier mode for the three highlighted boxes on the left using the same three colors for distinguishing the cases.   \label{fig:distiction}}
\end{figure*}

To clarify the distinction between the three types of solutions, in Fig. \ref{fig:distiction}.a we draw the spatiotemporal maps (not to scale, to save space) of the concentration profile of Homogeneous (red) , Turing (green ) and Mixed mode solutions (blue). In these maps, the differences between the solutions are not necessarily appreciable. To distinguish quantitatively the differences between those solutions, several factors were measured such as 1) the amplitude of the spatial pattern (the amplitude of the most predominant Fourier mode), 2) the way the predominant Fourier mode increased/decreased with growth/shrinkage, 3) the evolution of the wavenumber in the actual domain (not in the computational domain), and 4) the evolution in the phase plane of both zero-order Fourier mode concentrations. This study was performed for all the simulations, and the characteristic results of each type of solution (H, T or M) are exemplified in Fig. \ref{fig:distiction}.b. The homogeneous solutions have low amplitude, greater persistence of the predominant Fourier mode, and a constant steady state. The Turing pattern shares this last characteristic, but has a medium amplitude and a monotonic increase/decrease of the predominant  Fourier mode. The mixed mode shares this characteristic with the Turing pattern but has a larger amplitude and oscillates around a limit cycle.

\subsubsection{Amplitude and wavenumber of Turing patterns}

\begin{figure*}[hbtp]
\centering
\includegraphics[width=0.75 \textwidth]{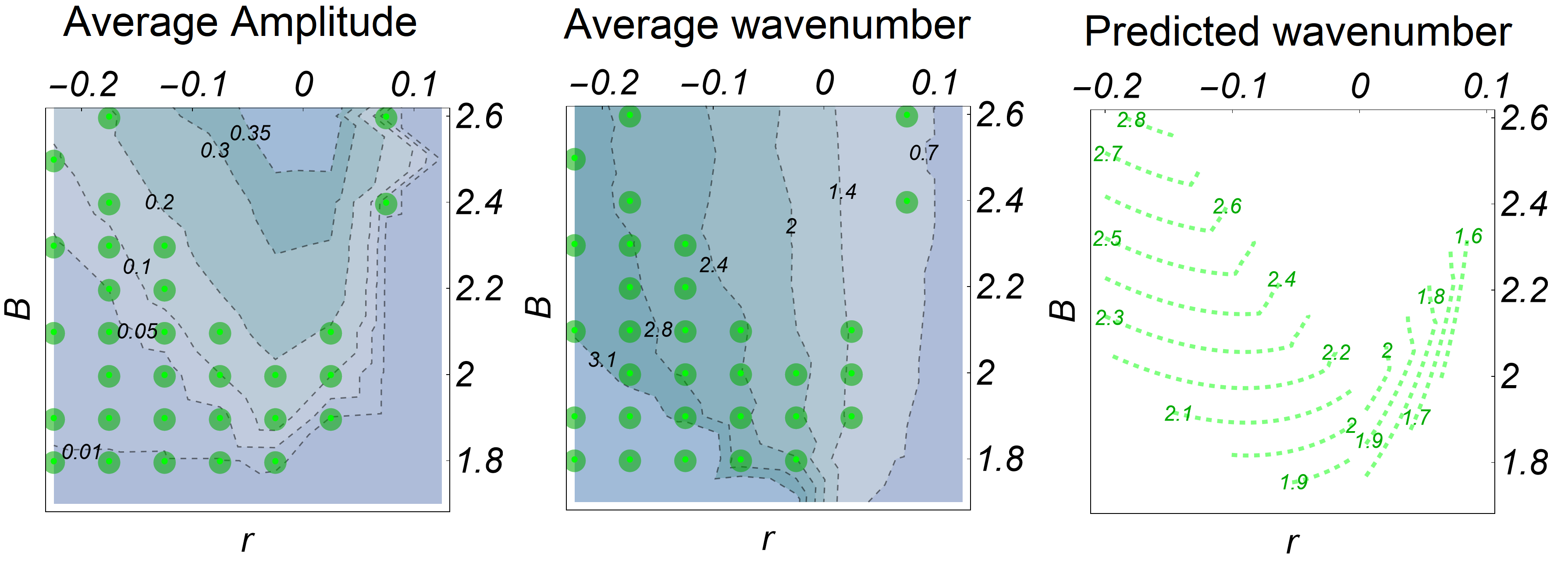}
\caption{Left and Center. Temporal averaged amplutide and wavenumber (in the  actual domain) of the numerical solutions. The Turing solutions are highligted in green. Right. Predicted wavenumber according to $k=k_u (r>0)$ or $k=k_v (r<0)$ given by our model. \label{fig:wavenumber}}
\end{figure*}

The three types of solutions presented so far are distinguished by their amplitude and wavenumber and have been previously exemplified in some cases. To have the complete picture, in Figure \ref{fig:wavenumber}.a and \ref{fig:wavenumber}.a we present the results of measuring the average amplitude and the wavenumber of all numerical solutions performed. As can be seen in \ref{fig:wavenumber}.a the average amplitudes grow from the Turing bifurcation (bottom of the region with the points illustrated in green) and continue to increase towards the region of mixed solutions. This agrees with what is known for Turing patterns in fixed domains, where the amplitude increases with the square root of the distance to the bifurcation \cite{ledesma2020eckhaus}. The aspect that this model does not predict correctly is related to the wavenumber which, as we illustrate in Fig. \ref{fig:wavenumber}.c, is expected to increase as you move away from the Turing bifurcation (as occurs in a fixed domain). As observed in the numerical result presented in Fig. \ref{fig:wavenumber}.b, the wave number depends more on the growth parameter $r$ than on the distance to the bifurcation (parabolic curve formed by the lower numerical points in green).

To carry out Fig. \ref{fig:wavenumber} we have taken into account that the instability condition is different depending on the sign of $r$; if $r>0$, the condition that destabilizes the system is I4 applied to the lowest diffusion coefficient, that is, the one related to $d_u$; for $r<0$, the bifurcation condition is the one related to $d_v$. From here we make the hypothesis that the wave number expressed for example in \eqref{eq:wavenumberc} must be $k_u$ and $k_v$, for $r$, positive and negative, respectively. Our model does predict correctly that for exponential growth the wavenumber in the average real domain remains essentially constant in time in general and, as confirmed by \ref{fig:wavenumber}, decreases as $r$ increases, however it does no t correctly predicts its value.

The reason for the discrepancy may be due to the fact that the wave number at a given moment crucially depends on its past history \cite{knobloch2015problems,ben2023turing}. This would mean that, for example, for $r>0$, at any time $t_1$, the pattern has a number of waves $N$; due to persistence \emph{i.e}, the ability of a system to preserve its wavenumber, the number of spatial waves will remain the same until it is outside the range of Eckhaus stability \cite{ledesma2020eckhaus}; at that moment, one more spatial wave (or perhaps several, depending on how fast the growth is) will enter to bring the system back within the range of $k$ values; unless the growth is very fast, we can assume that one more space wave will enter and the system will have $N+1$ waves, which will allow the solution to re-enter the stable $ k$ range; in a growing domain, this will generally make wave numbers in the low $k$ region more likely for growth. The opposite would happen on the left side ($r<0$), where the wavenumber tends to values greater than the one predicted in \ref{fig:wavenumber}.c, since, in general, the solution tends to gradually eliminate the waves that already had, making wave numbers at the top of the range more likely.

\begin{figure*}[hbtp]
\centering
\includegraphics[width=0.85 \textwidth ]{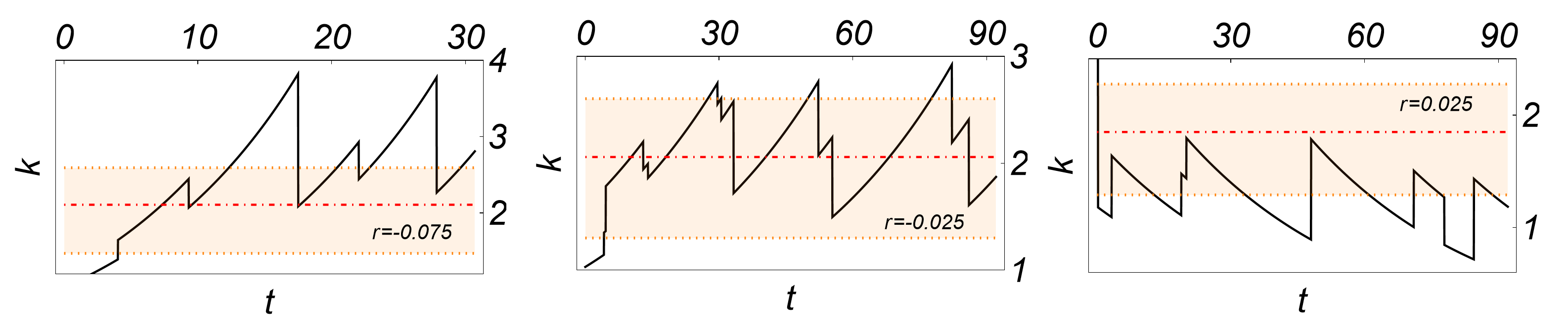}
\caption{Measured (solid black) and predicted wavenumber  (dotdashed red) and the range of posibble wavenumbers (between the orange dotted lines) for two cases of shrinkage and one of growing.  The growth rates $r$ are in the inset. \label{fig:range}}
\end{figure*}

To illustrate this wavenumber selection process, in Fig. \ref{fig:range}, we plot the wavenumber in the actual domain of the numerical solution (solid black lines), the expected wavenumber of the pattern (dashed line), according to $k=k_u (r>0)$ or $k=k_v (r<0)$, and the range of unstable wavenumbers $(k_{min},k_{max})=(\sqrt{k^2- \delta k^2},\sqrt{k^2+\delta k^2})$ (between the orange dotted lines) given by our model. For low values of $|r|$, the prediction of the range of wavenumbers is accurate and, as follows from the asymmetry to grow/shrink, for $r>0$, the wavenumbers lie at the bottom of the range, and the opposite occurs for $r<0$. However, as can be seen by comparing the two cases of shrinkage, the wavenumber range criterion fails for higher values of $|r|$ as the wavenumber of the numerical pattern lies outside the predicted range, probably due to that the inertial effects on pattern persistence are higher for more abrupt changes in the domain. However, this memory phenomenon will be studied in more detail in another Part of the work.

\section{Discussion and conclusions}\label{sec:discussion}

In this work we have proposed a new way of considering Turing destabilization for growing domains of two components. To do this, we have first considered the system of RDD equations for disturbances mapped to a fixed domain and have written them as a pair of second-order equations. We have rewritten each equation in such a way that each resembled the evolution of a system with a potential function. Hypothesizing that such a function would predict a destabilization of the same nature as it occurs in a fixed domain, we have generalized Turing's same ideas to an increasing domain by studying the deformation of such a function from a paraboloid to a saddle. We show that this strategy recovers the well-known Turing conditions for fixed domains and, for this Part of the work, we have exemplified it for exponential growth where the homogeneous state does not change considerably with time.

To demonstrate this, we use numerical simulations of the Brusselator and observe that the conditions predicted by our model allow us to give a good estimate of Turing space. These simulations show us that near the Turing region there are homogeneous solutions that show negligible amplitude and, at least numerically, tend to keep the same spatial mode. In addition, in the vicinity of the Hopf bifurcation, very stable spatial patterns appear with greater amplitude than Turing patterns, which differ from them only in that the homogeneous state is a stable limit cycle. The features of this type of solutions will be studied in another part of this work.

In this article, hypotheses are presented to give robust conditions for the formation of Turing patterns based on arguments about a possible potential function, contrasting with previous approximations whose comparison requires another work. However, our scheme allows us to understand the pattern formation process in a more general context related to the energy and entropy of dissipative structures. In this direction, in this work we have measured the Fourier spectrum of the solutions as a function of time and we have shown that the type of solution found (homogeneous, Turing or mixed) presents characteristics in observables such as the amplitude of the pattern and its average wavenumber. In the first case, we show that the Turing region is bound to the amplitude and that, therefore, the potential function found can represent an approximation to the energy of the system near the origin where the nonlinear terms would complete a kind of potential landscape \cite{ledesma2022energy}. 

Our scheme allows us to correctly predict that the wavenumber in actual real domain expressed by the system under exponential growth is maintained with oscillations around a constant value. Also, for growing/shrinking domain, the expressed wavenumber is generally less/greater than predicted, respectively, probably due to the fact that Turing patterns have some tendency to retain the wavenumber they had at the previous time, \emph{i.e.}, a type of memory \cite{krechetnikov2017stability,knobloch2014stability}. This issue of pattern persistence is related to the non-linearity of the  system and makes it difficult to determine the wavenumber the pattern will have based solely on a linear approximation and using a criterion, to put it in some way, static, that is, it does not depend on the initial or previous conditions of the system.

Despite these difficulties that arise mainly for large values of $|r|$, our scheme allows us to give closed analytic conditions for pattern formation for low values of $|r|$, which needless to say, are the most important from biological perspective, since in this case we can specify the fixed-poin concentrations, the Turing conditions, including the expected wave number and its range. In this sense, our approach presents a complete picture of this important process in diffusion reaction systems in both increasing and decreasing domains, the latter case being little studied in the literature.

Finally, it is worth mentioning that the detailed determination of the boundaries of the Turing region, especially for high values of $|r|$, requires much more detailed numerical simulations than those used in this work. This is because, for shrinkage, the lower bound of the region that distinguishes homogeneous solutions of Turing patterns leads to a very small characteristic time, so temporal refinement is necessary to make the average amplitude more representative. In the same way, for the upper edge, longer simulation times are required to be able to distinguish if the zero mode will actually tend to a limit cycle or decay to a constant asymptotic state. For the case of growth, since more spatial oscillations appear, considering longer times will require more spatial refinement to capture the evolution of the patterns. However, the precision used in this work is sufficient to conclude that our approximation very adequately captures the main features of the Turing space in growing domains.

\bibliography{biblio1}

\begin{thebibliography}{19}%
\makeatletter
\providecommand \@ifxundefined [1]{%
 \@ifx{#1\undefined}
}%
\providecommand \@ifnum [1]{%
 \ifnum #1\expandafter \@firstoftwo
 \else \expandafter \@secondoftwo
 \fi
}%
\providecommand \@ifx [1]{%
 \ifx #1\expandafter \@firstoftwo
 \else \expandafter \@secondoftwo
 \fi
}%
\providecommand \natexlab [1]{#1}%
\providecommand \enquote  [1]{``#1''}%
\providecommand \bibnamefont  [1]{#1}%
\providecommand \bibfnamefont [1]{#1}%
\providecommand \citenamefont [1]{#1}%
\providecommand \href@noop [0]{\@secondoftwo}%
\providecommand \href [0]{\begingroup \@sanitize@url \@href}%
\providecommand \@href[1]{\@@startlink{#1}\@@href}%
\providecommand \@@href[1]{\endgroup#1\@@endlink}%
\providecommand \@sanitize@url [0]{\catcode `\\12\catcode `\$12\catcode
  `\&12\catcode `\#12\catcode `\^12\catcode `\_12\catcode `\%12\relax}%
\providecommand \@@startlink[1]{}%
\providecommand \@@endlink[0]{}%
\providecommand \url  [0]{\begingroup\@sanitize@url \@url }%
\providecommand \@url [1]{\endgroup\@href {#1}{\urlprefix }}%
\providecommand \urlprefix  [0]{URL }%
\providecommand \Eprint [0]{\href }%
\providecommand \doibase [0]{http://dx.doi.org/}%
\providecommand \selectlanguage [0]{\@gobble}%
\providecommand \bibinfo  [0]{\@secondoftwo}%
\providecommand \bibfield  [0]{\@secondoftwo}%
\providecommand \translation [1]{[#1]}%
\providecommand \BibitemOpen [0]{}%
\providecommand \bibitemStop [0]{}%
\providecommand \bibitemNoStop [0]{.\EOS\space}%
\providecommand \EOS [0]{\spacefactor3000\relax}%
\providecommand \BibitemShut  [1]{\csname bibitem#1\endcsname}%
\let\auto@bib@innerbib\@empty
\bibitem [{\citenamefont {Krause}\ \emph {et~al.}(2019)\citenamefont {Krause},
  \citenamefont {Ellis},\ and\ \citenamefont
  {Van~Gorder}}]{krause2019influence}%
  \BibitemOpen
  \bibfield  {author} {\bibinfo {author} {\bibfnamefont {A.~L.}\ \bibnamefont
  {Krause}}, \bibinfo {author} {\bibfnamefont {M.~A.}\ \bibnamefont {Ellis}}, \
  and\ \bibinfo {author} {\bibfnamefont {R.~A.}\ \bibnamefont {Van~Gorder}},\
  }\href@noop {} {\bibfield  {journal} {\bibinfo  {journal} {Bulletin of
  mathematical biology}\ }\textbf {\bibinfo {volume} {81}},\ \bibinfo {pages}
  {759} (\bibinfo {year} {2019})}\BibitemShut {NoStop}%
\bibitem [{\citenamefont {Klika}\ and\ \citenamefont
  {Gaffney}(2017)}]{klika2017history}%
  \BibitemOpen
  \bibfield  {author} {\bibinfo {author} {\bibfnamefont {V.}~\bibnamefont
  {Klika}}\ and\ \bibinfo {author} {\bibfnamefont {E.~A.}\ \bibnamefont
  {Gaffney}},\ }\href@noop {} {\bibfield  {journal} {\bibinfo  {journal}
  {Proceedings of the Royal Society A: Mathematical, Physical and Engineering
  Sciences}\ }\textbf {\bibinfo {volume} {473}},\ \bibinfo {pages} {20160744}
  (\bibinfo {year} {2017})}\BibitemShut {NoStop}%
\bibitem [{\citenamefont {Ledesma-Dur{\'a}n}\ \emph {et~al.}(2020)\citenamefont
  {Ledesma-Dur{\'a}n}, \citenamefont {Ortiz-Dur{\'a}n}, \citenamefont
  {Arag{\'o}n},\ and\ \citenamefont
  {Santamar{\'\i}a-Holek}}]{ledesma2020eckhaus}%
  \BibitemOpen
  \bibfield  {author} {\bibinfo {author} {\bibfnamefont {A.}~\bibnamefont
  {Ledesma-Dur{\'a}n}}, \bibinfo {author} {\bibfnamefont {E.}~\bibnamefont
  {Ortiz-Dur{\'a}n}}, \bibinfo {author} {\bibfnamefont {J.}~\bibnamefont
  {Arag{\'o}n}}, \ and\ \bibinfo {author} {\bibfnamefont {I.}~\bibnamefont
  {Santamar{\'\i}a-Holek}},\ }\href@noop {} {\bibfield  {journal} {\bibinfo
  {journal} {Physical Review E}\ }\textbf {\bibinfo {volume} {102}},\ \bibinfo
  {pages} {032214} (\bibinfo {year} {2020})}\BibitemShut {NoStop}%
\bibitem [{\citenamefont {Madzvamuse}\ \emph {et~al.}(2010)\citenamefont
  {Madzvamuse}, \citenamefont {Gaffney},\ and\ \citenamefont
  {Maini}}]{madzvamuse2010stability}%
  \BibitemOpen
  \bibfield  {author} {\bibinfo {author} {\bibfnamefont {A.}~\bibnamefont
  {Madzvamuse}}, \bibinfo {author} {\bibfnamefont {E.~A.}\ \bibnamefont
  {Gaffney}}, \ and\ \bibinfo {author} {\bibfnamefont {P.~K.}\ \bibnamefont
  {Maini}},\ }\href@noop {} {\bibfield  {journal} {\bibinfo  {journal} {Journal
  of mathematical biology}\ }\textbf {\bibinfo {volume} {61}},\ \bibinfo
  {pages} {133} (\bibinfo {year} {2010})}\BibitemShut {NoStop}%
\bibitem [{\citenamefont {Van~Gorder}\ \emph {et~al.}(2021)\citenamefont
  {Van~Gorder}, \citenamefont {Klika},\ and\ \citenamefont
  {Krause}}]{van2021turing}%
  \BibitemOpen
  \bibfield  {author} {\bibinfo {author} {\bibfnamefont {R.~A.}\ \bibnamefont
  {Van~Gorder}}, \bibinfo {author} {\bibfnamefont {V.}~\bibnamefont {Klika}}, \
  and\ \bibinfo {author} {\bibfnamefont {A.~L.}\ \bibnamefont {Krause}},\
  }\href@noop {} {\bibfield  {journal} {\bibinfo  {journal} {Journal of
  mathematical biology}\ }\textbf {\bibinfo {volume} {82}},\ \bibinfo {pages}
  {1} (\bibinfo {year} {2021})}\BibitemShut {NoStop}%
\bibitem [{\citenamefont {Crampin}\ \emph {et~al.}(1999)\citenamefont
  {Crampin}, \citenamefont {Gaffney},\ and\ \citenamefont
  {Maini}}]{crampin1999reaction}%
  \BibitemOpen
  \bibfield  {author} {\bibinfo {author} {\bibfnamefont {E.~J.}\ \bibnamefont
  {Crampin}}, \bibinfo {author} {\bibfnamefont {E.~A.}\ \bibnamefont
  {Gaffney}}, \ and\ \bibinfo {author} {\bibfnamefont {P.~K.}\ \bibnamefont
  {Maini}},\ }\href@noop {} {\bibfield  {journal} {\bibinfo  {journal}
  {Bulletin of mathematical biology}\ }\textbf {\bibinfo {volume} {61}},\
  \bibinfo {pages} {1093} (\bibinfo {year} {1999})}\BibitemShut {NoStop}%
\bibitem [{\citenamefont {Pe\~{n}a}\ and\ \citenamefont
  {Perez-Garcia}(2001)}]{pena2001stability}%
  \BibitemOpen
  \bibfield  {author} {\bibinfo {author} {\bibfnamefont {B.}~\bibnamefont
  {Pe\~{n}a}}\ and\ \bibinfo {author} {\bibfnamefont {C.}~\bibnamefont
  {Perez-Garcia}},\ }\href@noop {} {\bibfield  {journal} {\bibinfo  {journal}
  {Physical review E}\ }\textbf {\bibinfo {volume} {64}},\ \bibinfo {pages}
  {056213} (\bibinfo {year} {2001})}\BibitemShut {NoStop}%
\bibitem [{\citenamefont {Lepp{\"a}nen}\ \emph {et~al.}(2004)\citenamefont
  {Lepp{\"a}nen} \emph {et~al.}}]{leppanen2004computational}%
  \BibitemOpen
  \bibfield  {author} {\bibinfo {author} {\bibfnamefont {T.}~\bibnamefont
  {Lepp{\"a}nen}} \emph {et~al.},\ }\href@noop {} {\emph {\bibinfo {title}
  {Computational studies of pattern formation in Turing systems}}}\ (\bibinfo
  {publisher} {Helsinki University of Technology},\ \bibinfo {year}
  {2004})\BibitemShut {NoStop}%
\bibitem [{\citenamefont {Toole}\ and\ \citenamefont
  {Hurdal}(2013)}]{toole2013turing}%
  \BibitemOpen
  \bibfield  {author} {\bibinfo {author} {\bibfnamefont {G.}~\bibnamefont
  {Toole}}\ and\ \bibinfo {author} {\bibfnamefont {M.~K.}\ \bibnamefont
  {Hurdal}},\ }\href@noop {} {\bibfield  {journal} {\bibinfo  {journal}
  {Computers \& Mathematics with Applications}\ }\textbf {\bibinfo {volume}
  {66}},\ \bibinfo {pages} {1627} (\bibinfo {year} {2013})}\BibitemShut
  {NoStop}%
\bibitem [{\citenamefont {Murray}(2001)}]{murray2001mathematical}%
  \BibitemOpen
  \bibfield  {author} {\bibinfo {author} {\bibfnamefont {J.~D.}\ \bibnamefont
  {Murray}},\ }\href@noop {} {\emph {\bibinfo {title} {Mathematical biology II:
  Spatial models and biomedical applications}}},\ Vol.~\bibinfo {volume} {3}\
  (\bibinfo  {publisher} {Springer New York},\ \bibinfo {year}
  {2001})\BibitemShut {NoStop}%
\bibitem [{\citenamefont {Ledesma-Dur{\'a}n}\ and\ \citenamefont
  {Santamar{\'\i}a-Holek}(2022)}]{ledesma2022energy}%
  \BibitemOpen
  \bibfield  {author} {\bibinfo {author} {\bibfnamefont {A.}~\bibnamefont
  {Ledesma-Dur{\'a}n}}\ and\ \bibinfo {author} {\bibfnamefont {I.}~\bibnamefont
  {Santamar{\'\i}a-Holek}},\ }\href@noop {} {\bibfield  {journal} {\bibinfo
  {journal} {Journal of Non-Equilibrium Thermodynamics}\ }\textbf {\bibinfo
  {volume} {47}},\ \bibinfo {pages} {311} (\bibinfo {year} {2022})}\BibitemShut
  {NoStop}%
\bibitem [{\citenamefont {Gjorgjieva}\ and\ \citenamefont
  {Jacobsen}(2007)}]{gjorgjieva2007turing}%
  \BibitemOpen
  \bibfield  {author} {\bibinfo {author} {\bibfnamefont {J.}~\bibnamefont
  {Gjorgjieva}}\ and\ \bibinfo {author} {\bibfnamefont {J.}~\bibnamefont
  {Jacobsen}},\ }in\ \href@noop {} {\emph {\bibinfo {booktitle} {Conference
  Publications}}},\ Vol.\ \bibinfo {volume} {2007}\ (\bibinfo {organization}
  {Conference Publications},\ \bibinfo {year} {2007})\ pp.\ \bibinfo {pages}
  {436--445}\BibitemShut {NoStop}%
\bibitem [{\citenamefont {Challenger}\ \emph {et~al.}(2015)\citenamefont
  {Challenger}, \citenamefont {Burioni},\ and\ \citenamefont
  {Fanelli}}]{challenger2015turing}%
  \BibitemOpen
  \bibfield  {author} {\bibinfo {author} {\bibfnamefont {J.~D.}\ \bibnamefont
  {Challenger}}, \bibinfo {author} {\bibfnamefont {R.}~\bibnamefont {Burioni}},
  \ and\ \bibinfo {author} {\bibfnamefont {D.}~\bibnamefont {Fanelli}},\
  }\href@noop {} {\bibfield  {journal} {\bibinfo  {journal} {Physical Review
  E}\ }\textbf {\bibinfo {volume} {92}},\ \bibinfo {pages} {022818} (\bibinfo
  {year} {2015})}\BibitemShut {NoStop}%
\bibitem [{\citenamefont {Ledesma-Dur{\'a}n}\ and\ \citenamefont
  {Arag{\'o}n}(2019)}]{ledesma2019primary}%
  \BibitemOpen
  \bibfield  {author} {\bibinfo {author} {\bibfnamefont {A.}~\bibnamefont
  {Ledesma-Dur{\'a}n}}\ and\ \bibinfo {author} {\bibfnamefont {J.~L.}\
  \bibnamefont {Arag{\'o}n}},\ }\href@noop {} {\bibfield  {journal} {\bibinfo
  {journal} {Chaos, Solitons \& Fractals}\ }\textbf {\bibinfo {volume} {124}},\
  \bibinfo {pages} {68} (\bibinfo {year} {2019})}\BibitemShut {NoStop}%
\bibitem [{\citenamefont {Arag{\'o}n}\ \emph {et~al.}(2012)\citenamefont
  {Arag{\'o}n}, \citenamefont {Barrio}, \citenamefont {Woolley}, \citenamefont
  {Baker},\ and\ \citenamefont {Maini}}]{aragon2012nonlinear}%
  \BibitemOpen
  \bibfield  {author} {\bibinfo {author} {\bibfnamefont {J.}~\bibnamefont
  {Arag{\'o}n}}, \bibinfo {author} {\bibfnamefont {R.}~\bibnamefont {Barrio}},
  \bibinfo {author} {\bibfnamefont {T.}~\bibnamefont {Woolley}}, \bibinfo
  {author} {\bibfnamefont {R.}~\bibnamefont {Baker}}, \ and\ \bibinfo {author}
  {\bibfnamefont {P.}~\bibnamefont {Maini}},\ }\href@noop {} {\bibfield
  {journal} {\bibinfo  {journal} {Physical Review E}\ }\textbf {\bibinfo
  {volume} {86}},\ \bibinfo {pages} {026201} (\bibinfo {year}
  {2012})}\BibitemShut {NoStop}%
\bibitem [{\citenamefont {Knobloch}\ and\ \citenamefont
  {Krechetnikov}(2015)}]{knobloch2015problems}%
  \BibitemOpen
  \bibfield  {author} {\bibinfo {author} {\bibfnamefont {E.}~\bibnamefont
  {Knobloch}}\ and\ \bibinfo {author} {\bibfnamefont {R.}~\bibnamefont
  {Krechetnikov}},\ }\href@noop {} {\bibfield  {journal} {\bibinfo  {journal}
  {Acta Applicandae Mathematicae}\ }\textbf {\bibinfo {volume} {137}},\
  \bibinfo {pages} {123} (\bibinfo {year} {2015})}\BibitemShut {NoStop}%
\bibitem [{\citenamefont {Ben~Tahar}\ \emph {et~al.}(2023)\citenamefont
  {Ben~Tahar}, \citenamefont {Mu{\~n}oz}, \citenamefont {Shefelbine},\ and\
  \citenamefont {Comellas}}]{ben2023turing}%
  \BibitemOpen
  \bibfield  {author} {\bibinfo {author} {\bibfnamefont {S.}~\bibnamefont
  {Ben~Tahar}}, \bibinfo {author} {\bibfnamefont {J.~J.}\ \bibnamefont
  {Mu{\~n}oz}}, \bibinfo {author} {\bibfnamefont {S.~J.}\ \bibnamefont
  {Shefelbine}}, \ and\ \bibinfo {author} {\bibfnamefont {E.}~\bibnamefont
  {Comellas}},\ }\href@noop {} {\bibfield  {journal} {\bibinfo  {journal}
  {bioRxiv}\ ,\ \bibinfo {pages} {2023}} (\bibinfo {year} {2023})}\BibitemShut
  {NoStop}%
\bibitem [{\citenamefont {Krechetnikov}\ and\ \citenamefont
  {Knobloch}(2017)}]{krechetnikov2017stability}%
  \BibitemOpen
  \bibfield  {author} {\bibinfo {author} {\bibfnamefont {R.}~\bibnamefont
  {Krechetnikov}}\ and\ \bibinfo {author} {\bibfnamefont {E.}~\bibnamefont
  {Knobloch}},\ }\href@noop {} {\bibfield  {journal} {\bibinfo  {journal}
  {Physica D: Nonlinear Phenomena}\ }\textbf {\bibinfo {volume} {342}},\
  \bibinfo {pages} {16} (\bibinfo {year} {2017})}\BibitemShut {NoStop}%
\bibitem [{\citenamefont {Knobloch}\ and\ \citenamefont
  {Krechetnikov}(2014)}]{knobloch2014stability}%
  \BibitemOpen
  \bibfield  {author} {\bibinfo {author} {\bibfnamefont {E.}~\bibnamefont
  {Knobloch}}\ and\ \bibinfo {author} {\bibfnamefont {R.}~\bibnamefont
  {Krechetnikov}},\ }\href@noop {} {\bibfield  {journal} {\bibinfo  {journal}
  {Journal of Nonlinear Science}\ }\textbf {\bibinfo {volume} {24}},\ \bibinfo
  {pages} {493} (\bibinfo {year} {2014})}\BibitemShut {NoStop}%
\end{thebibliography}%

\end{document}